\documentclass[11pt,a4paper]{article} 
\usepackage{times}
\usepackage{bbm, amsmath, amsthm, amssymb}       
\usepackage{mathrsfs}
\usepackage[dvips]{color, graphicx} 
\usepackage{picinpar}
\usepackage[below]{placeins}
\usepackage{subfig}
\newcommand{\eV}{\, \textrm{eV}}

\newcommand{\GeV}{\, \textrm{GeV}}

\begin{document}
\bibliographystyle{h-physrev3}

\title{ 
%
\bf Effects of reheating on leptogenesis\\[8mm]}
\author{F.~Hahn-Woernle and M.~Pl\"umacher\\[1ex]
Max Planck Institute for Physics, F\"ohringer Ring 6, 80805 Munich, Germany}
\date{}
\maketitle

\thispagestyle{empty}

\begin{abstract}
  We study the evolution of a cosmological baryon asymmetry in 
  leptogenesis when the right-handed neutrinos are produced in
  inflaton decays. By performing a detailed numerical study over
  a broad range of inflaton-neutrino couplings we show that the
  resulting asymmetry can be larger by two orders of magnitude or
  more than in thermal leptogenesis, if the reheating temperature
  $T_{RH}$ is of the same order as the right-handed neutrino mass
  $M_1$. Hence, the lower limit on the baryogenesis temperature
  obtained in thermal leptogenesis can be relaxed accordingly.
\end{abstract}

\section{Introduction}
\label{sec:introduction}

The observed cosmological baryon asymmetry can naturally be explained
via decays of heavy right-handed neutrinos (RHN), a
scenario known as leptogenesis~\cite{Fukugita:1986hr}. In its simplest
form, thermal leptogenesis, the baryon asymmetry is produced during
the radiation dominated era and stringent limits on neutrino
parameters are obtained. In particular, succesful leptogenesis
requires the reheating temperature of the universe to be larger than
about $4\times10^8\,$GeV. In the favoured strong-washout regime an
even stricter lower limit on $T_{\rm RH}$ of about $4\times10^9\,$GeV
is obtained. This may be in conflict with big bang nucleosynthesis
(BBN) in supergravity (SUGRA) models due to the gravitino
problem. There successful BBN is only possible if $T_{\rm RH}$ is
lower than about $10^{6-7}\,$GeV\cite{Kawasaki:2004qu}.

An alternative production mechanism is considered in non-thermal
leptogenesis
models~\cite{Lazarides:1991wu,Murayama:1992ua,Kumekawa:1994gx,Kolb:1996jt,Giudice:1999fb,
  Asaka:1999yd,Asaka:1999jb,Jeannerot:2001qu,Hamaguchi:2001gw,Giudice:2003jh}
where one assumes that right-handed neutrinos are produced directly in
the decays of some heavier particle. That particle could be the
inflaton, the particle related to an inflationary phase in the very
early universe. Supersymmetric~\cite{Murayama:1992ua,Kumekawa:1994gx,
  Asaka:1999yd,Asaka:1999jb,Giudice:1999fb,Jeannerot:2001qu,Hamaguchi:2001gw}
and grand unified models~\cite{Lazarides:1991wu} have been considered,
the focus of all these studies being put on the underlying model
of inflation in order to derive the coupling between the inflaton and the
right-handed neutrino\footnote{There are recent attempts to couple the
  inflaton to the neutrino in a non direct way to allow for instant
  non-thermal leptogenesis~\cite{Endo:2006nj}}. Most of these models
have in common that the decay width of the inflaton,
$\Gamma_{\Phi}$, is much smaller than the decay width of the
neutrino, $\Gamma_{N}$, i.e.\ $\Gamma_{\Phi} \ll \Gamma_{N}$. Hence, the
neutrino decay instantaneously follows the inflaton decay and the reheating
temperature $T_{RH}$ is much smaller than the RHN mass $M_1$. In such 
scenarios the produced baryon asymmetry can easily be evaluated without
the need for a full numerical investigation.
 In this
work we will also consider the case $T_{RH} \sim M_{1}$ and show that here
a full numerical study by means of Boltzmann equations is needed.
We will discuss quantitatively the dependence of the final baryon asymmetry on
$\Gamma_{\Phi}$ and $\Gamma_{N}$ for a broad range of parameters.
Furthermore, we will see how the bounds on neutrino parameters and the
reheating temperature derived
in thermal leptogenesis are relaxed.


In the next section we briefly review thermal leptogenesis in order to
set the scene and introduce some notation that will be used in the rest
of the paper. In section \ref{sec:lg_i_decay} we introduce our model for
the inflaton-neutrino coupling and discuss the case that the decay width
of the inflaton is much smaller than that of the neutrino. If that is not
the case, a more detailed treatment in terms of Boltzmann equations is
needed, which are introduced in section \ref{sec:fund_BE}. Finally,
in section \ref{sec:results} we present and discuss our results and their
dependence on the parameters of the inflaton and the neutrino.

\section{Thermal Leptogenesis}
\label{sec:thermal_lg}
In the standard thermal leptogenesis scenario the right-handed
neutrinos are produced dynamically by scattering processes in the
thermal bath or are assumed to be initially in thermal equilibrium.
Usually a hierarchical mass scheme is assumed, $M_{3},M_{2} \gg
M_{1}$. Then the lightest right-handed neutrino, $N_{1}$, decays into
a standard model lepton-Higgs pair, $N_1 \rightarrow H+l_{L}$ and 
$N_1 \rightarrow H^{\dagger} + l_{L}^{\dagger}$ generating a lepton 
asymmetry if $CP$ is not conserved in the decay.  The
decay rate of $N_1$ reads~\cite{Buchmuller:2003gz}
\begin{equation}
  \label{eq:2}
   \Gamma_{N}=
   H(M_{1})\, K \, \frac{K_{1}(z)}{K_{2}(z)},
\end{equation}
where $K_{1}, K_{2}$ are Bessel functions and $H$ is the Hubble
parameter. The parameter $K$,
\begin{equation}
  \label{eq:3}
    K \equiv \frac{\Gamma_{N}(z=\infty)}{H(z=1)} = \frac{\tilde{m}_{1}}{m_{\ast}},
\end{equation}
separates the regions of weak washout, $K\ll1$ and strong washout, $K\gg1$.
Here, $\tilde{m_{1}}$ is the effective light neutrino mass and
$m_{\ast} \simeq 1.1 \times 10^{-3} \eV$. $\Gamma(z=\infty)$ is the
decay rate in the rest frame of the particle, i.e.\ the
decay width $\Gamma_{N_{rf}}= H(M_{1})K$.

The maximal CP asymmetry in the decays of $N_1$ is given
by~\cite{Buchmuller:2002jk}\cite{Covi:1997dr}:
\begin{equation}
  \label{eq:6}
  \epsilon^{\textrm{max}}_{1} = \frac{3}{16 \pi} \frac{M_{1}m_{3}}
  {v^{2}}\beta \approx 10^{-6} \left( \frac{M_{1}}{10^{10} \GeV}
  \right) \left( \frac{m_{3}}{0.05 \eV} \right)\,\beta,
\end{equation}
where $\beta \leq 1$, with the exact value depending on different
see-saw parameters.
This maximal CP asymmetry then yields the maximal baryon asymmetry
$\eta^{\textrm{max}}_{B}$ that can be produced in leptogenesis. 
Since $\epsilon_{1}^{\textrm{max}}\propto M_1$, requiring that the
maximal baryon asymmetry is larger that the observed one, i.e.\
\mbox{$\eta^{max}_{B} \geq \eta^{CMB}_{B}$}, yields a lower bound
on $M_1$~\cite{Buchmuller:2004nz},
\begin{equation}
  \label{eq:M_1_bound}
  \begin{aligned}
    M_{1} > M^{\textrm{min}}_{1} &= \frac{1}{d} \frac{16 \pi}{3}
    \frac{v^{2}} {m_{\textrm{atm}}} \frac{\eta^{CMB}_{B}}{\kappa_{f}}
    \\ &\approx 6.4 \times 10^{8} \; \textrm{GeV}
    \left(\frac{\eta^{CMB}_{B}}{6 \times 10^{10}} \right) \left(\frac{
        0.05 \; \textrm{eV}}{m_{\textrm{atm}}} \right)
    \kappa^{-1}_{f}.
  \end{aligned}
\end{equation}
Here $\kappa_{f}$ is the final efficiency factor which parametrizes
the dynamics of the lepton asymmetry and neutrino
production~\cite{Buchmuller:2005eh}. It is obtained by solving the
relevant set of Boltzmann equations.
 In thermal leptogenesis its
maximum value is by definition $1$ for the case of a thermal initial
abundance of right-handed neutrinos. 
In the factor \mbox{$d = 3 \alpha_{\textrm{sph}} / (4f)
  \approx 0.96 \times 10^{-2}$}, $f=2387/86$ accounts for the dilution due to
photon production from the onset of leptogenesis till recombination  and
the factor $\alpha_{\textrm{sph}} =28/79$ accounts for the partial conversion
of the lepton asymmetry into a baryon asymmetry by sphaleron processes.

Since all this is assumed to happen in the thermal, radiation dominated phase
of the universe, the lower bound on $M_{1}$ translates into a lower bound on
the initial temperature of leptogenesis, which corresponds to a lower limit on
the reheating temperature after inflation~\cite{Buchmuller:2005eh},
\begin{equation}
  \label{eq:12}
  T_{RH} \approx M^{\textrm{min}}_{1} \geq 4 \times 10^{8} \, {\rm GeV}\;.
\end{equation}
Such a large reheating temperature is potentially in conflict with
BBN in supersymmetric models, where upper bounds on the reheating 
temperature as low as $10^6\,$GeV have been obtained in SUGRA
models~\cite{Kawasaki:2004qu}. This and the dependence of the
produced baryon asymmetry on the initial conditions in the weak
washout regime are some of the shortcomings of thermal leptogenesis.
Hence, it is worthwhile to study alternative leptogenesis scenarios.
In the following we will discuss a scenario where the neutrinos
are produced non-thermally in inflaton decays. As we shall see,
the lower bound on the reheating temperature can be relaxed by
as much as three orders of magnitude in the most interesting parameter
range of strong washout.

\section{Leptogenesis via inflaton decay}
\label{sec:lg_i_decay}
In the following we will assume that the inflaton $\Phi$ decays
exclusively into a
pair of the lightest right-handed neutrinos, $\Phi \rightarrow
N_{1}+N_{1}$. 
The decay width for this process can be
parametrized as
\begin{equation}
  \label{eq:1}
  \Gamma_{\Phi} \simeq \frac{\vert \gamma \vert^{2}}{4 \pi} M_{\Phi}\;,
\end{equation}
$\gamma$ being the inflaton-neutrino coupling.
Further, we assume a hierarchical mass spectrum for the heavy
neutrinos, $M_{3},M_{2} \gg M_{1}$, hence potential effects of $N_{2}$ and
$N_{3}$ can be neglected\footnote{Note, however, that the baryon asymmetry
may also be generated by the second-lightest heavy neutrino in certain
areas of parameter space \cite{DiBari:2005st}.}.

Neglecting potential contributions from preheating 
\cite{Giudice:1999fb,GarciaBellido:2001cb}, which are
generically rather small anyway \cite{Allahverdi:1996xc,Prokopec:1996rr}, 
the decay considered above
is kinematically allowed if $M_{\Phi} \geq 2 M_{1}$, which will
always be the case in the following.

After the inflaton condensate has decayed away, the  heavy neutrinos
dominate the energy density of the universe, a scenario
known as dominant initial abundance. When the
right-handed neutrinos have become non-relativistic they decay in the 
standard way, thereby producing a lepton asymmetry, and reheat the 
universe since their decay products, standard
model lepton and Higgs doublets, quickly thermalize.

The reheating temperature is usually computed assuming that
the energy stored in the inflaton condensate is
instantaneously transformed into radiation. This yields
\begin{equation}
  \label{eq:4}
  T_{RH} = \left(\frac{90}{8\pi^{3}g^{\ast}} \right)^{\frac{1}{4}}
  \sqrt{\Gamma_{\Phi}M_{Pl}} = 0.06 \; \vert \gamma \vert \, \left(\frac{200}{g^{\ast}}
  \right)^{\frac{1}{4}} \sqrt{M_{\Phi}M_{Pl}} ,
\end{equation}
where $g^{\ast}$ is the number of relativistic degrees of freedom at
$T_{RH}$. Analogously, one can define a reheating temperature for the
reheating process due to neutrino decays:
\begin{equation}
  \label{eq:5}
  T_{RH}^{N} = \left(\frac{90}{8\pi^{3}g^{\ast}} \right)^{\frac{1}{4}} 
  \sqrt{\Gamma_{N_{rf}} M_{Pl}} = M_{1}\; \sqrt{K}.
\end{equation}
In the decay chain considered the real physical reheating temperature
is given by Eq.~(\ref{eq:5}), since only after the neutrinos have
decayed is the thermal bath of the radiation dominated universe
produced.
$T_{RH}$ from Eq.~(\ref{eq:4}) will
be used to parametrize the inflaton-neutrino coupling. Only in models
where $\Gamma_{\Phi} \ll \Gamma_{N}$, does $T_{RH}$ correspond to the real
physical reheating temperature since then the neutrino mass $M_{1}$ is much
larger than $T_{RH}$ and, therefore, the right-handed neutrinos decay
instantaneously after having been produced in inflaton decays.
The resulting lepton asymmetry in
such scenarios can easily be evaluated \cite{Lazarides:1991wu,Murayama:1992ua}.
After reheating the baryon asymmetry, defined here
as the ratio of bayon number to photon density, is given by:
\begin{equation}
  \label{eq:7}
  \frac{n_B}{n_{\gamma}} =  \alpha_{sph} \epsilon_{1}
  \frac{n_{N_{1}}}{n_{\gamma}} = 
  \alpha_{sph} \epsilon_{1} \frac{T_{RH}}{30M_{1}} 
  \simeq 10^{-8} \frac{T_{RH}}{M_{1}}.
\end{equation}
Here, we have set $\kappa=1$ since washout processes are completely
negligible in this case. Furthermore, we have assumed that the energy
density of the heavy neutrino is instantaneously converted into
relativistic degrees of freedom, yielding a temperature for the
thermal bath $\rho_{R}=\rho_{N_{1}}= (\pi^{2}/30)g_{\ast}T^{4}$.
Again demanding that the baryon asymmetry is at least equal to the
observed value one gets a constraint on the reheating temperature
\begin{equation}
  \label{eq:8}
  T_{RH} \geq 10^{-2} M_{1}\;,
\end{equation}
which corresponds to a lower limit on the inflaton-neutrino coupling
$\gamma$ through Eq.~(\ref{eq:4}).

\section{The Boltzmann equations}
\label{sec:fund_BE}
In this work we shall consider a more general range of parameters.
In particular we shall discuss in detail the case when the reheating
temperature is of the same order as the heavy neutrino mass. Then,
the simple approximation discussed above does not hold anymore,
and the asymmetry has to be computed by solving a system of
Boltzmann equations.
We shall study in detail the dependence of the final efficiency
factor $\kappa_f$ on the inflaton-neutrino coupling, again parametrized
by the reheating temperature $T_{RH}$.
We will see that in this model there is a strong correlation between
the  reheating temperature and the neutrino mass via the decay widths
$\Gamma_{N_{rf}}$ and $\Gamma_{\Phi}$. 

The relevant Boltzmann equations for the energy densities of the
inflaton, the lightest of the heavy right-handed neutrinos, the $B-L$
asymmetry and the radiation energy density, respectively, read as
follows\footnote{Note that we are neglecting an inverse decay term
  $\sim \Gamma_{\Phi} \rho_{\Phi}^{eq}$ since the reheating
  temperature is assumed to be much smaller than the inflaton mass and
  hence the inflaton never comes into thermal equilibrium.}:
\begin{equation}
  \label{eq:9}
   \begin{aligned}
    \dot{\rho}_{\Phi} &= -3H\rho_{\Phi}-\Gamma_{\Phi}\,\rho_{\Phi} \\
    \dot{\rho}_{N} &= -3H\rho_{N}+\Gamma_{\Phi}\,\rho_{\Phi}-\Gamma_{N}(\rho_{N}-\rho_{N}^{eq}) \\
    \dot{n}_{B-L} &= -3Hn_{B-L}-\epsilon\Gamma_{N}(n_{N}-n_{N}^{eq})
    -\Gamma_{ID}\, n_{B-L} \\
    \dot\rho_{R} &= -4H\rho_{R}+\Gamma_{N}(\rho_{N}-\rho_{N}^{eq})\;.
  \end{aligned}
\end{equation}
Here we consider only decays and inverse decays and neglect scattering
processes of the right-handed neutrinos and the inflaton. Further, in
the Boltzmann equation for $\rho_N$ we assumed that the right-handed
neutrinos are non-relativistic.
Surely this is an
assumption which is not guaranteed and, depending on the mass of the
inflaton, the produced heavy neutrinos can have energies much larger
than their rest mass. But, this approximation works well for the
computation of $\kappa_{f}$, which is our main purpose.  The reason is
that for all values of $K$ the final asymmetry is determined by
right-handed neutrino decays when they are fully non-relativistic.
Therefore the ultra-relativistic stage is not important for the final
efficiency factor. An exact treatment would be relevant for the
description of the evolution of the universe in the interval between
inflation and the decay of the right-handed neutrinos as well as for
the computation of the maximal value of the temperature,
$T_{\textrm{max}}$, after inflation. As is shown
in~\cite{Chung:1998rq}, $T_{\textrm{max}}$ can be much larger than the
reheating temperature. However, this goes beyond the scope of this
investigation.

In the actual numerical integration of these equations it is useful
to use quantities in which the expansion of the universe has been
scaled out. The relevant variables as well as the transformed Boltzmann
equations and some numerical parameters are discussed in Appendix
A. For definiteness we will always assume a neutrino mass
$M_1=10^9\,$GeV and an inflaton mass $M_{\Phi}=10^{13}\,$GeV and only
vary the reheating temperature, i.e.\ the inflaton-neutrino coupling,
and $K$ in the following.

\section{Results and Discussion} 
\label{sec:results}

In this section we shall present our results for the final efficiency
factor $\kappa_f$, as shown in Fig.~\ref{fig:non_thermal_kf_M9Tall}.
In particular, we will discuss the dependence on the reheating
temperature $T_{RH}$, i.e.\ the inflaton-neutrino coupling, and
$K$.

\subsection{General Observations}

\begin{figure}[t]
  \centering
  \includegraphics[width=0.7\textwidth]{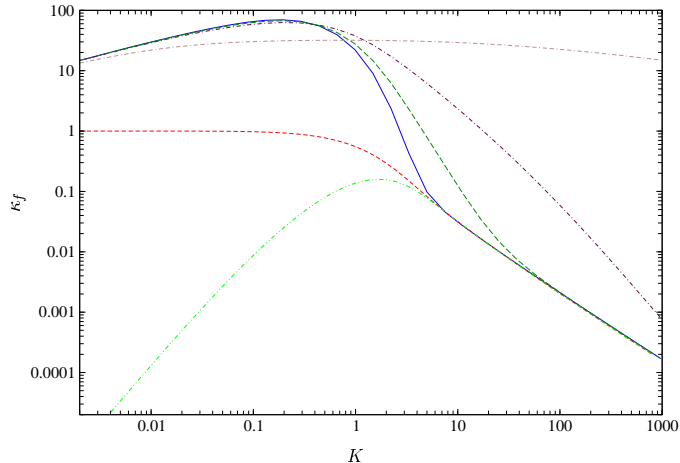}
  \caption{\small {The final efficiency for thermal (short-dashed), zero
    (point-point-dashed) and dominant initial $N_{1}$ abundance for
    $T_{RH}=10^{9}$ GeV (solid), $T_{RH}=5 \times 10^{8}$ Gev
    (long-dashed), $T_{RH}=3.75 \times 10^{8}$ GeV (point-dashed and
    $T_{RH}=10^{8}$ GeV (point-dashed-dashed))}}
  \label{fig:non_thermal_kf_M9Tall}
\end{figure}

In Fig.~\ref{fig:non_thermal_kf_M9Tall} we have plotted our results for the
final efficiency factor as a function of $K$ for various values of the
reheating temperature as well as the standard results obtained in thermal
leptogenesis~\cite{Buchmuller:2004nz}. As one can see, for
$T_{RH}\geq5\times10^8\,$GeV our results are in good agreement with the ones
obtained in thermal leptogenesis in the strong washout regime, as one would
naively expect. On the other hand, in the weak washout regime $\kappa_f$ is
enhanced by up to two orders of magnitude. Further, the curves for $T_{RH}
\geq 10^{9}\GeV$ are in agreement with the results obtained
in~\cite{Giudice:2003jh}, where values $T_{RH} \leq M_{1}$ were not
considered.


In the strong washout regime, i.e.\ for $K>1$, the neutrino Yukawa
coupling is large enough to keep the system close to equilibrium, 
thereby erasing any dependence on the initial conditions, as long
as the physical reheating temperature is of order of the neutrino
mass or larger, so that the right-handed neutrinos still can 
thermalize.



For
reheating temperatures above $T_{RH}=10^{9} \GeV$ $\kappa_{f}$ for
dominant initial $N_{1}$ abundance is in good agreement with those for
zero and thermal initial abundance in the whole strong washout regime
for $K \gtrsim 5$. When the rehating temperature becomes smaller than
the neutrino mass, e.g.\ for $T_{RH}=5 \times 10^{8}
\GeV$ one sees an agreement only for very strong washout, $K
\geq 40$. For $T_{RH}= 3.75 \times 10^{8}$ GeV, the washout effects
are so weak that the final efficiency factor is larger than the one
for thermal and zero initial $N_{1}$ abundance for all values of
$K$.

In the weak washout regime, on the other hand, one can see that $\kappa_f$
for dominant initial $N_1$ abundance is quite independent of $T_{RH}$
and much larger than one, which is by definition the largest value for
thermal initial $N_1$ abundance. This is due to the direct production
of neutrinos via inflaton decays which leads to neutrino abundances
larger than the equilibrium value.


\subsection{Weak Washout Regime}

In the weak washout regime, i.e.\ $K<1$,  $\kappa_{f}$ is much larger
than in thermal leptogenesis, by a factor $\sim 10-100$ for all considered
reheating temperatures and is almost independent of the reheating
temperature. The physical reheating temperature, given by 
$T_{RH}^{N} = M_{1} \sqrt{K}$, is smaller than the
right-handed neutrino mass in the whole weak washout regime. Hence,
the neutrinos decay strongly out of equilibrium.

The maximum value of $\kappa_{f}$ is reached at $ K \sim 0.4$ and is
almost independent of $T_{RH}$. This is due to the fact that the
entropy produced in each neutrino decay, $\Delta S \sim M_{1}/T_{RH}^{N}$,
corresponding to an increase of the number of photons, 
$\Delta N_{\gamma} \propto \Delta S$, becomes larger at small $K$, 
since the neutrinos decay later and hence carry a greater fraction
of the energy density of the universe when they decay.

\begin{figure}[t!]
  \label{fig:all_M19T19_K0_01}
  \FloatBarrier 
  \subfloat[]{
    \includegraphics[width=0.5\textwidth]{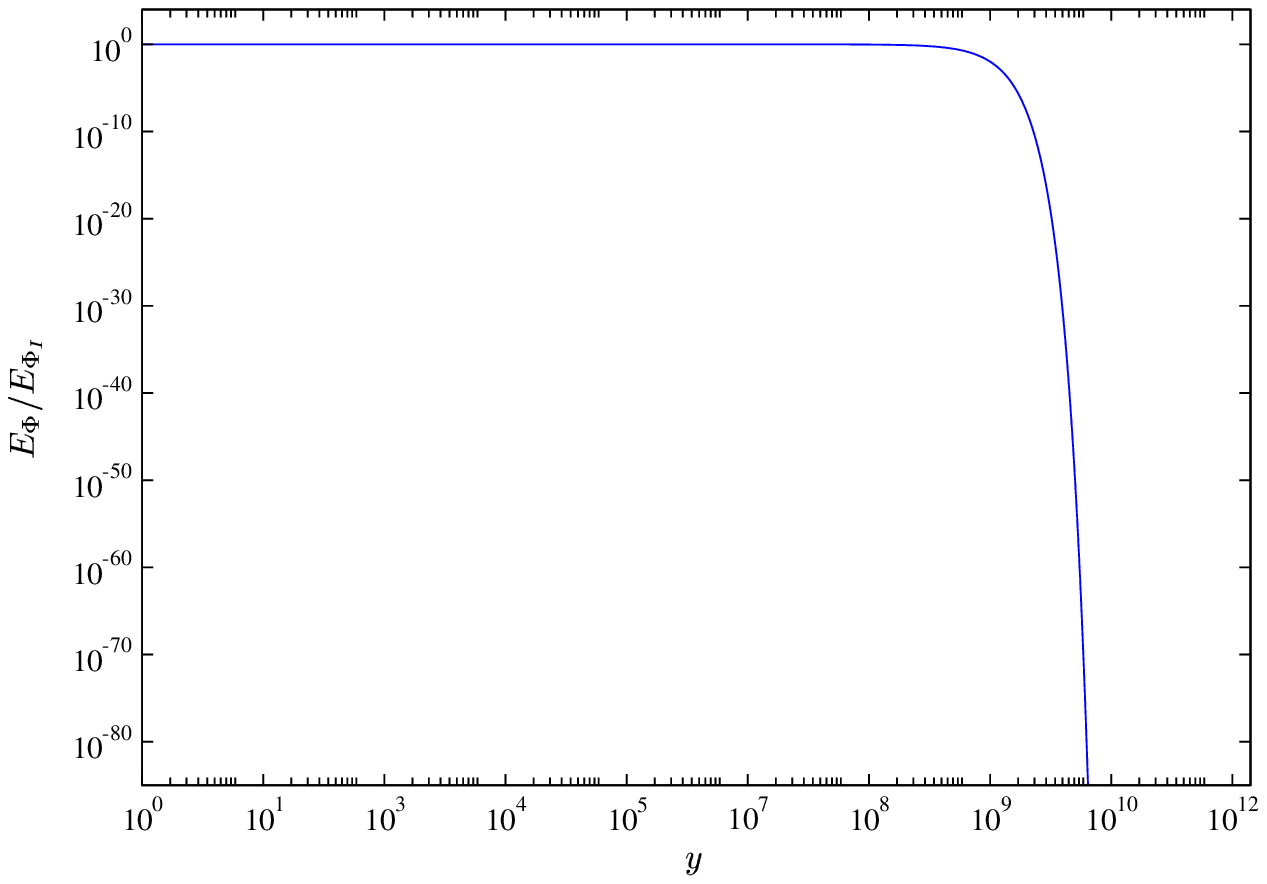}
    \label{fig:Fi_M19T19_K0_01}} 
  \subfloat[]{
    \includegraphics[width=0.5\textwidth]{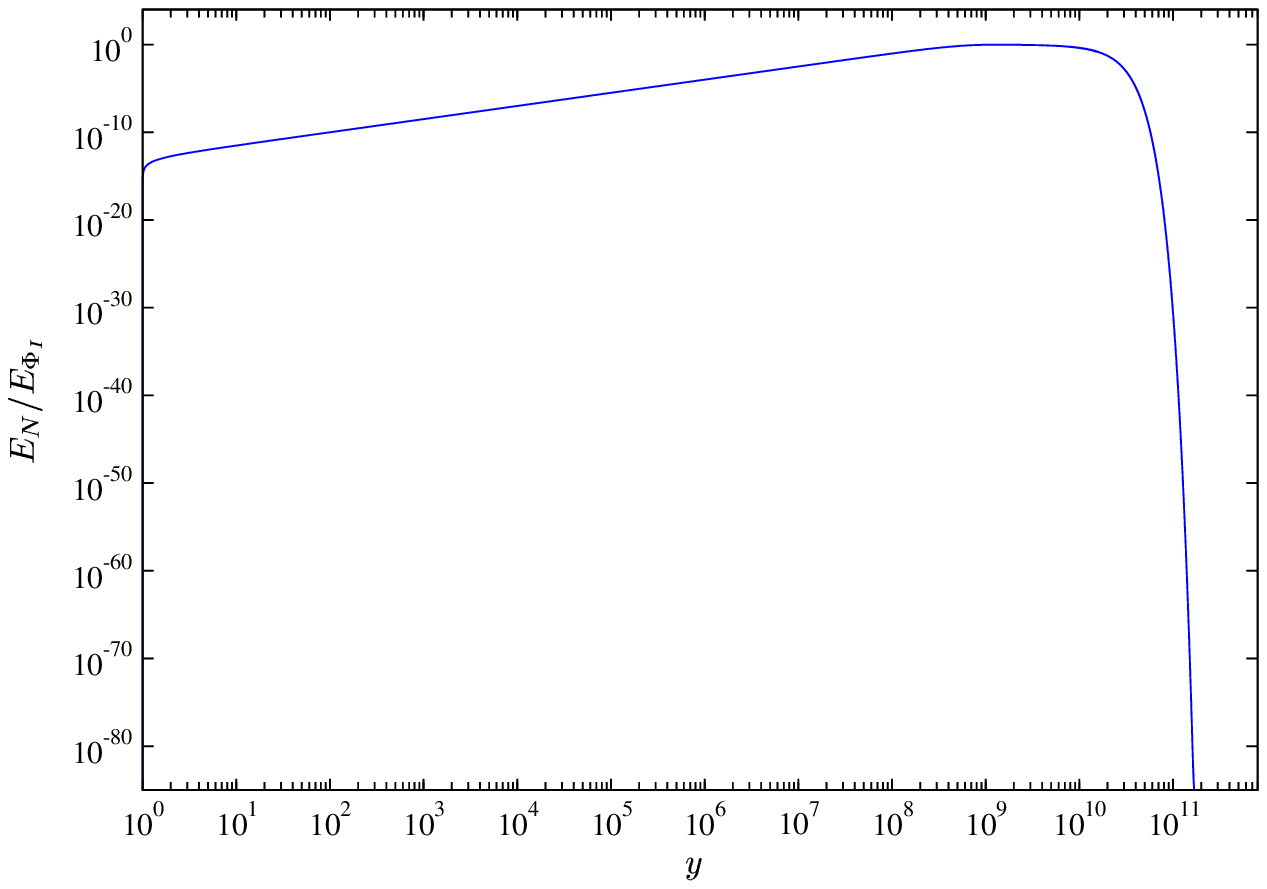}
    \label{fig:N_M19T19_K0_01}}
  \subfloat[]{
    \includegraphics[width=0.5\textwidth]{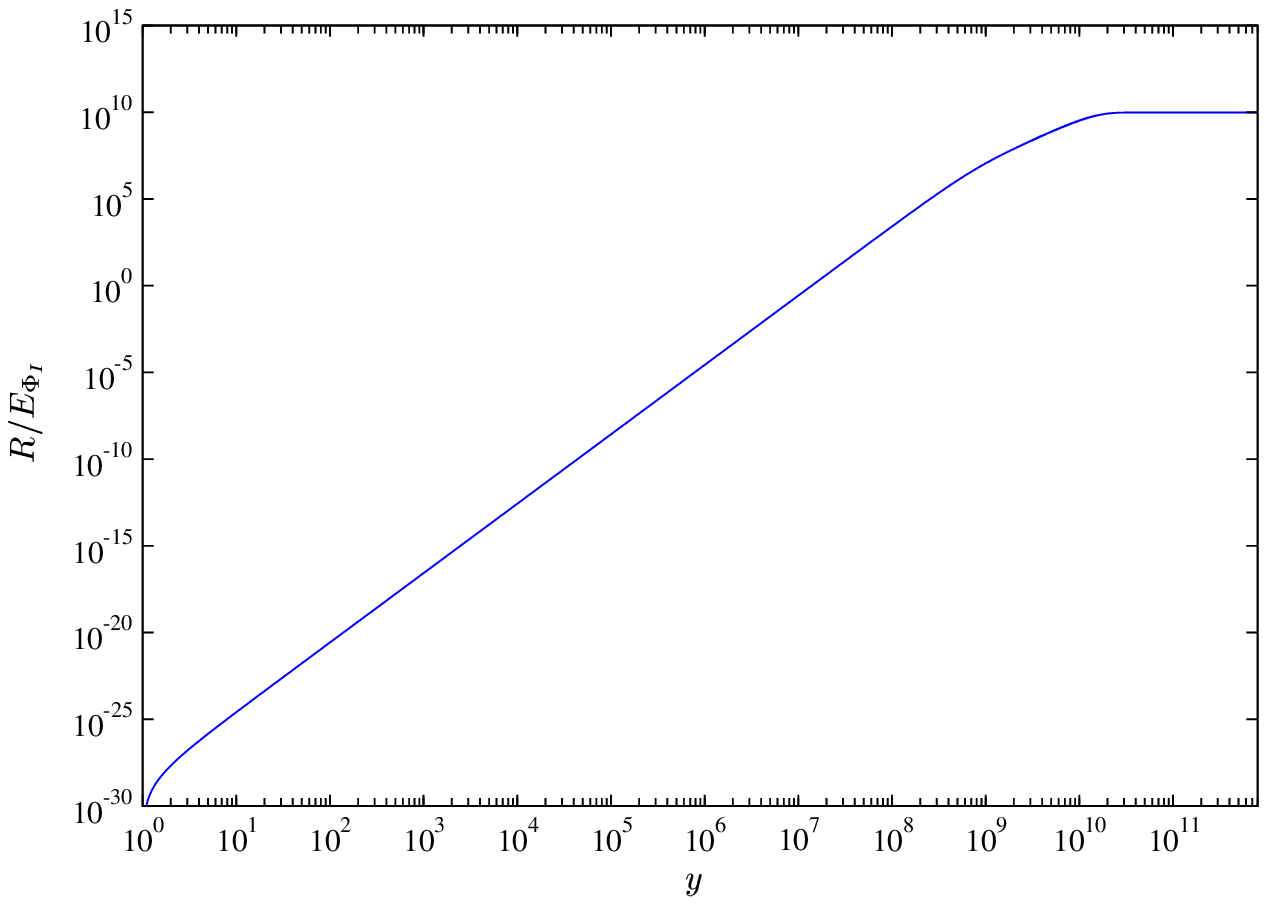}
    \label{fig:R_M19T19_K0_01}}
  \subfloat[]{
    \includegraphics[width=0.5\textwidth]{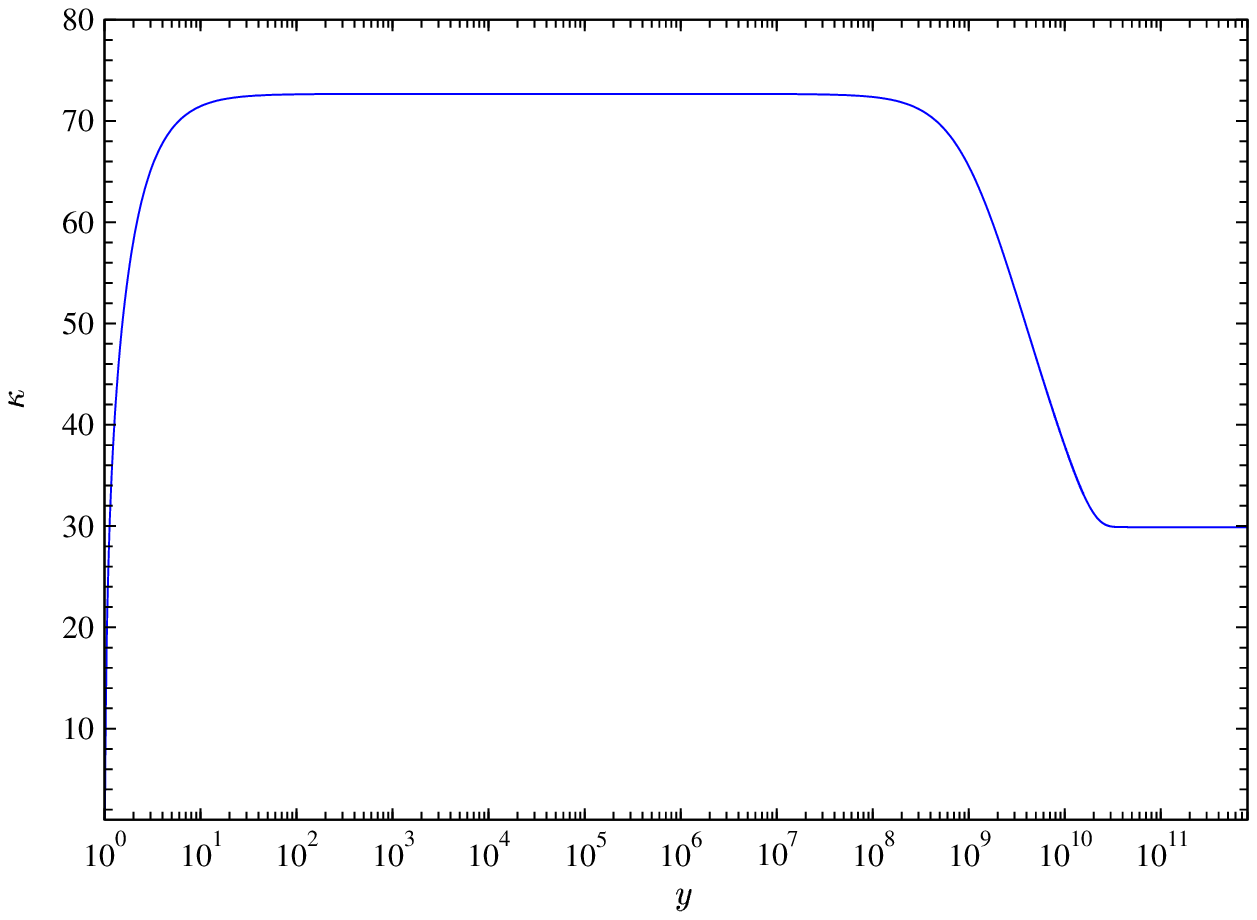}
    \label{fig:k_M19T19_K0_01}}
    \subfloat[]{
      \includegraphics[width=0.5\textwidth]{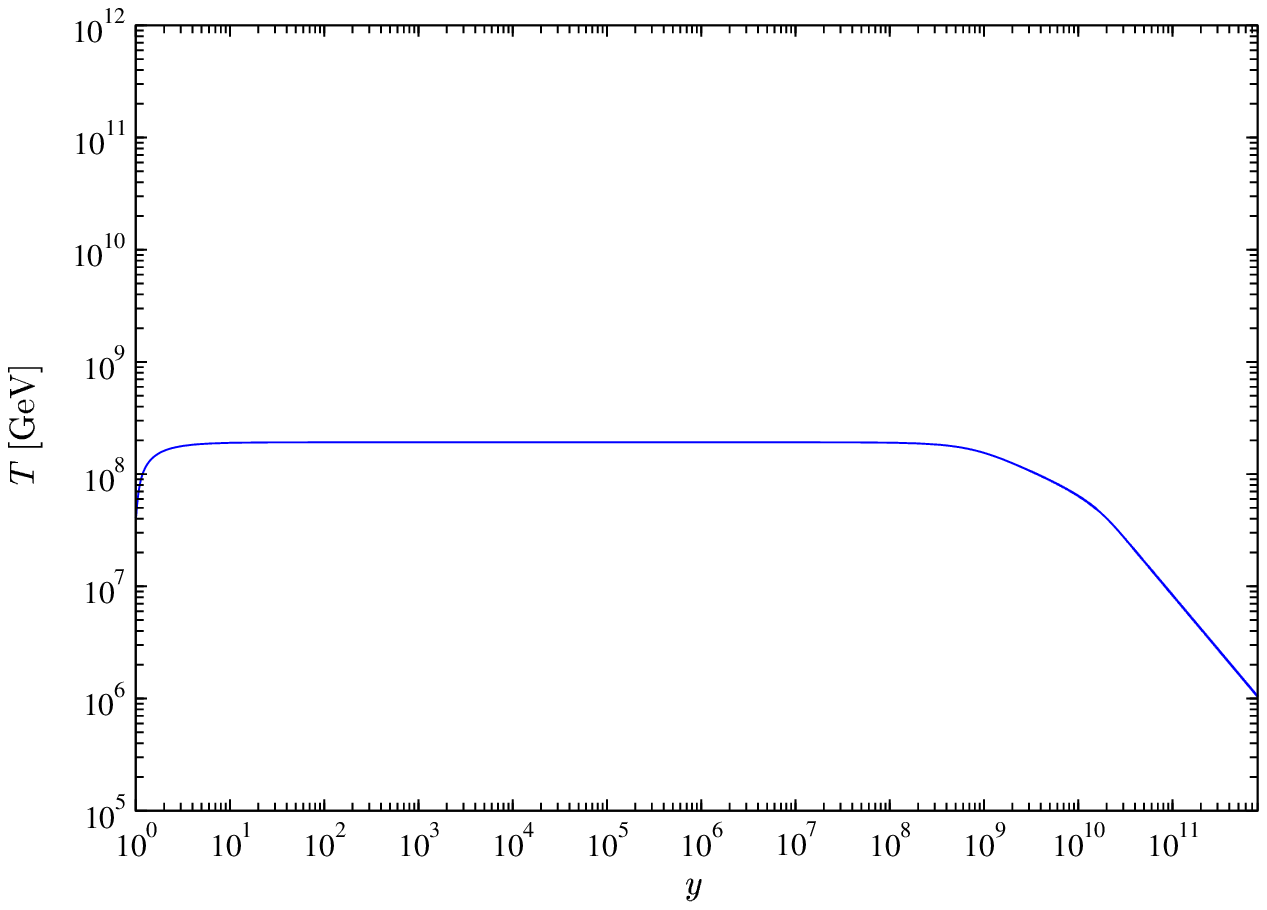}
      \label{fig:T_M19T19_K0_01}}
    \subfloat[]{
      \includegraphics[width=0.5\textwidth]{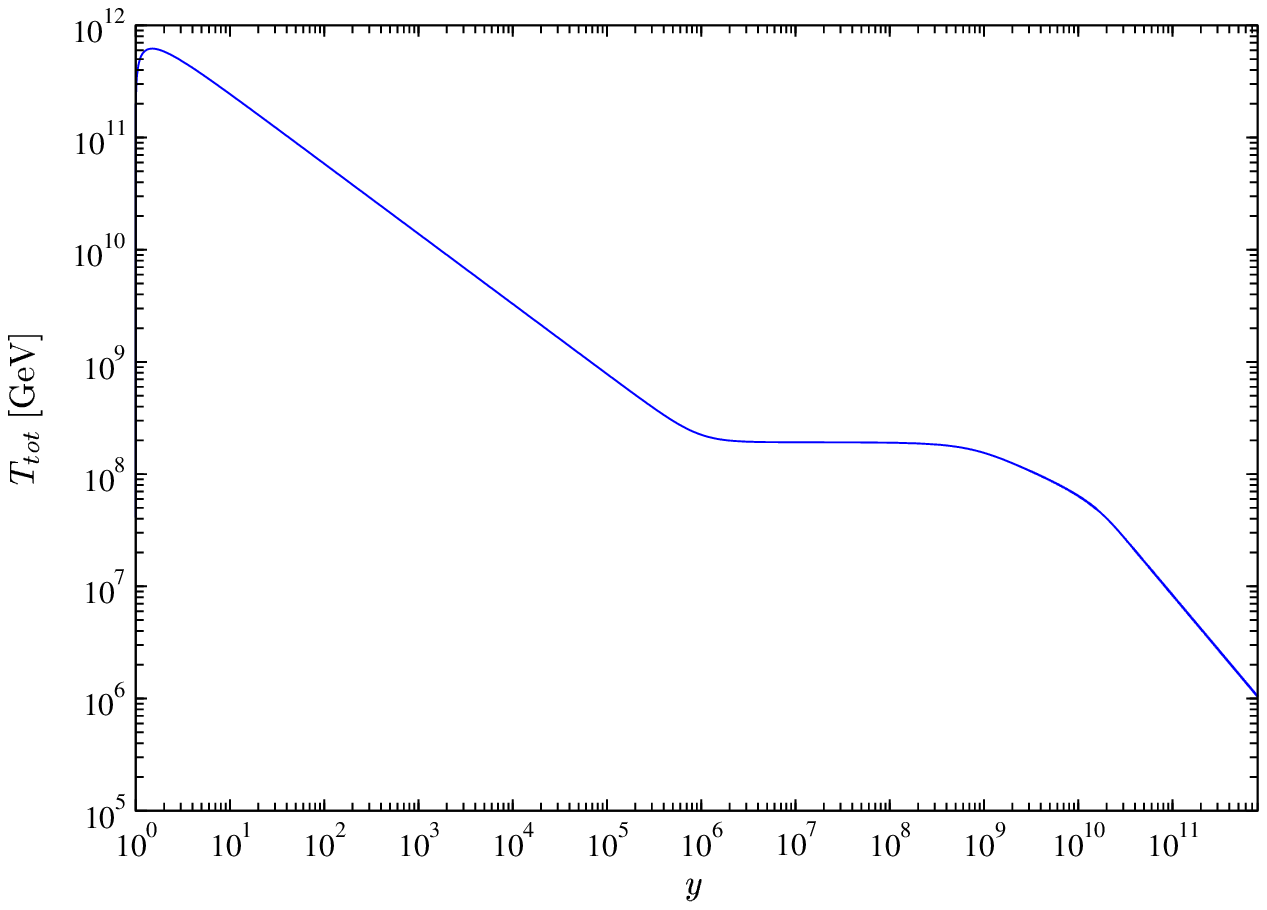}
      \label{fig:T-tot_M19T19_K0_01}}
    \caption{\small{The evolution of $E_{\Phi}/E_{\Phi_{I}}$,
      $E_{N}/E_{\Phi_{I}}$, $R$ $\kappa$, $T$ and $T_{tot}$ is
      shown for $T_{RH}=10^{9}$ GeV at $K=10^{-2}$}}
\end{figure}

In figure~\ref{fig:Fi_M19T19_K0_01} one can see that for $T_{RH}=10^9\,$GeV
and $K=10^{-2}$
the inflaton starts decaying when the scale factor $y$, defined in
Appendix A, reaches $\sim 10^9$ while at $y\sim 10^{10}$
it has completely decayed. Figure~\ref{fig:N_M19T19_K0_01} shows
that when the inflaton starts decaying at $y=10^9$ the energy density
of the heavy neutrinos becomes constant and dominates the energy
density of the universe. At $y \sim 5 \times 10^{10}$ the heavy
neutrinos start decaying and the associated entropy release triggers a
transition from a matter dominated to a radiation dominated universe
as one can deduce from figure~\ref{fig:R_M19T19_K0_01} where the
rescaled radiation energy density becomes constant. The evolution of the
efficiency factor is plotted in figure~\ref{fig:k_M19T19_K0_01}. It
remains constant at a value $\kappa \sim 80$ as long as the universe
is (matter) dominated by the inflaton and gets reduced when the
universe is $N_{1}$ (matter) dominated. After the neutrinos have
decayed the efficiency factor reaches its final value of about $\kappa
\sim 30$. In figures~\ref{fig:T_M19T19_K0_01} and
~\ref{fig:T-tot_M19T19_K0_01} we show the evolution of the
temperature. The difference in the two plots is that in
figure~\ref{fig:T_M19T19_K0_01} we have taken into account only the
contribution from radiation,
\begin{equation}
\label{eq:21}
  T = \left[\frac{30\rho_{R}}{\pi^{2}g^{\ast}} \right]^{\frac{1}{4}}. 
\end{equation}
This gives the temperature as long as the heavy neutrinos, produced in
the inflaton decays, are non-relativistic. When $M_{\Phi} \gg M_{1}$
the produced right-handed neutrinos are relativistic particles and
their energy density contributes to the energy density of radiation
that determines the temperature. This effect is included in
figure~\ref{fig:T-tot_M19T19_K0_01}. Here, we have assumed that all the
produced heavy neutrinos are relativistic particles and defined the
temperature as:
\begin{equation}
  \label{eq:22}
    T = \left[\frac{30(\rho_{R}+\rho_{N})}{\pi^{2}g^{\ast}} \right]^{\frac{1}{4}}. 
\end{equation}
This, of course, is a rather rough approximation, since in the Boltzmann
equations, Eqs.~(\ref{eq:9}), the right-handed
neutrinos are treated as non-relativistic particles. However, this shows 
that, cf.\ figure~\ref{fig:T-tot_M19T19_K0_01}, the maximal
temperature achieved in the reheating process is much larger than the
physical reheating temperature\cite{Chung:1998rq}, which in this case is
$T_{RH}^{N} \sim 7 \times 10^{7}\GeV$, as can be read off
directly from figure~\ref{fig:T_M19T19_K0_01}.\\

\subsection{Strong  Washout Regime}

In the strong washout regime, i.e.\ $K>1$, the final efficiency factor
$\kappa_f$ is in perfect agreement with the results obtained in thermal
leptogenesis, as long as $T_{RH}\gtrsim M_1$. This is what one would
expect, since all reactions involving $N_1$ are in thermal equilibrium,
hence the neutrinos rapidly thermalize and any information about the
initial conditions is quickly lost. For reheating temperatures smaller
than $M_1$ this is not necessarily the case anymore, e.g.\ for 
$T_{RH}=5\times10^8\,$GeV and $K\lesssim50$, the reactions involving
$N_1$ are not strong enough to bring them into thermal equilibrium,
i.e.\ the neutrinos decay rather strongly out of equilibrium. Hence,
the final efficiency factor is enhanced compared to thermal leptogenesis
since washout processes are suppressed.

\begin{figure}[t!]
  \label{fig:all_M19T19_K500} 
  \FloatBarrier
  \subfloat[]{
  \includegraphics[width=0.5\textwidth]{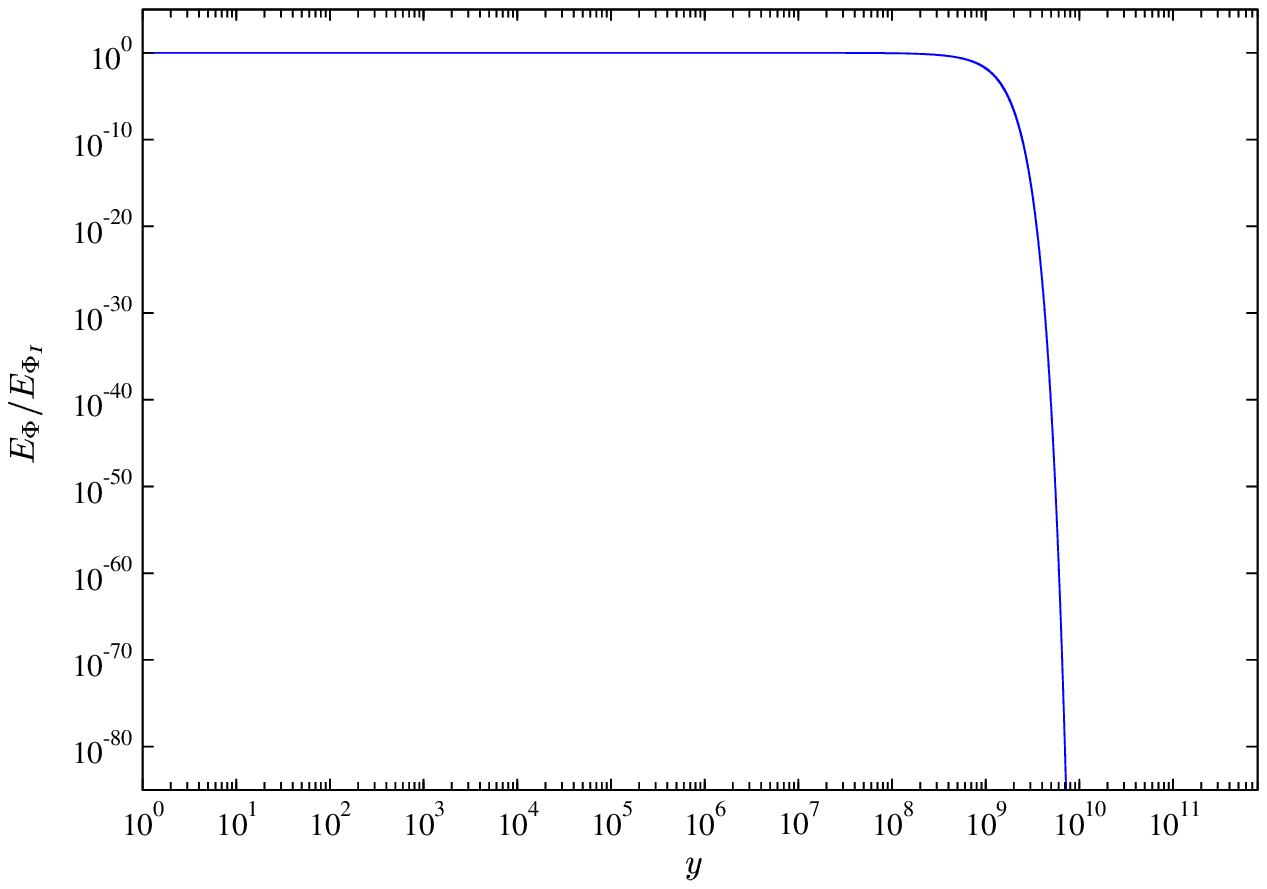}  
   \label{fig:Fi_M19T19_K500}} 
   \subfloat[]{
  \includegraphics[width=0.5\textwidth]{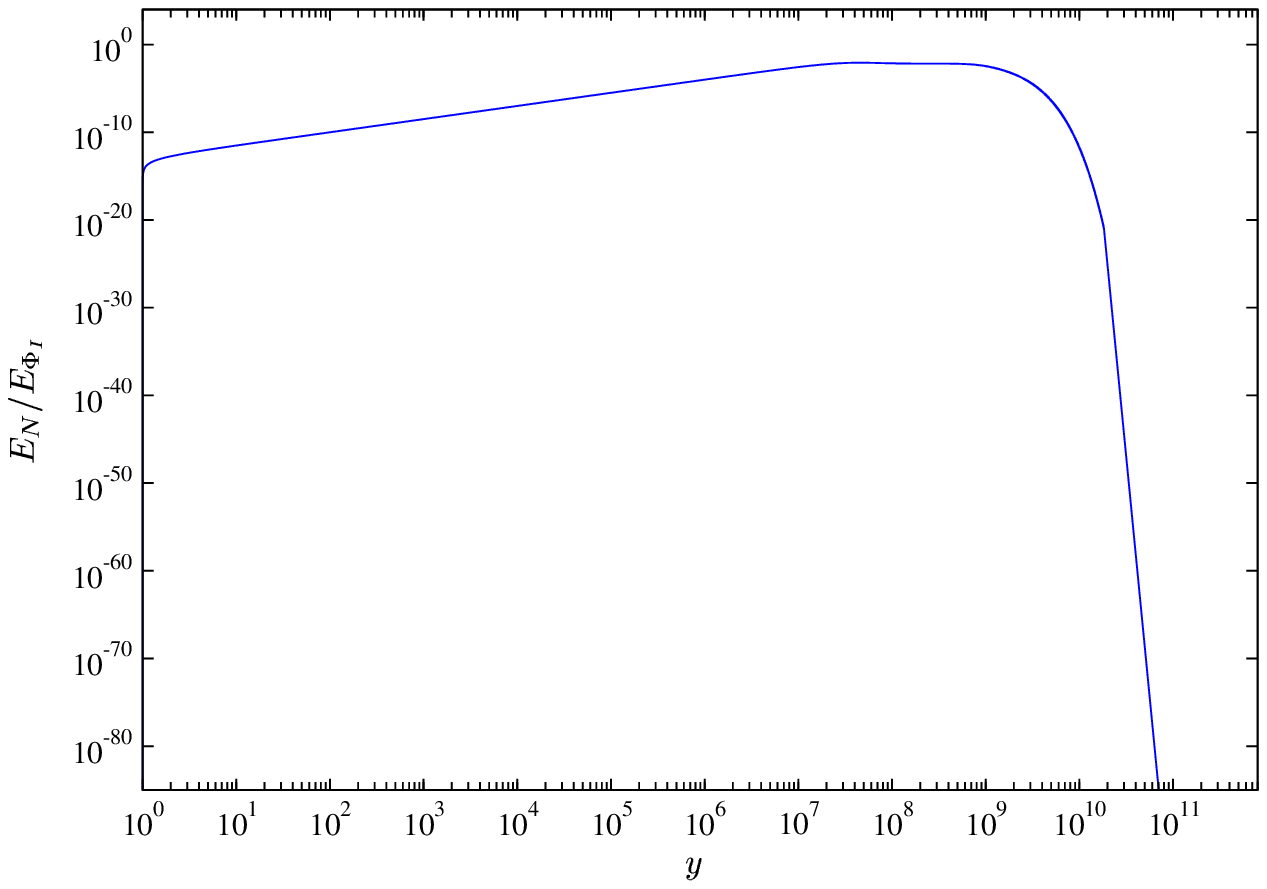}
    \label{fig:N_M19T19_K500}}
   \subfloat[]{
  \includegraphics[width=0.5\textwidth]{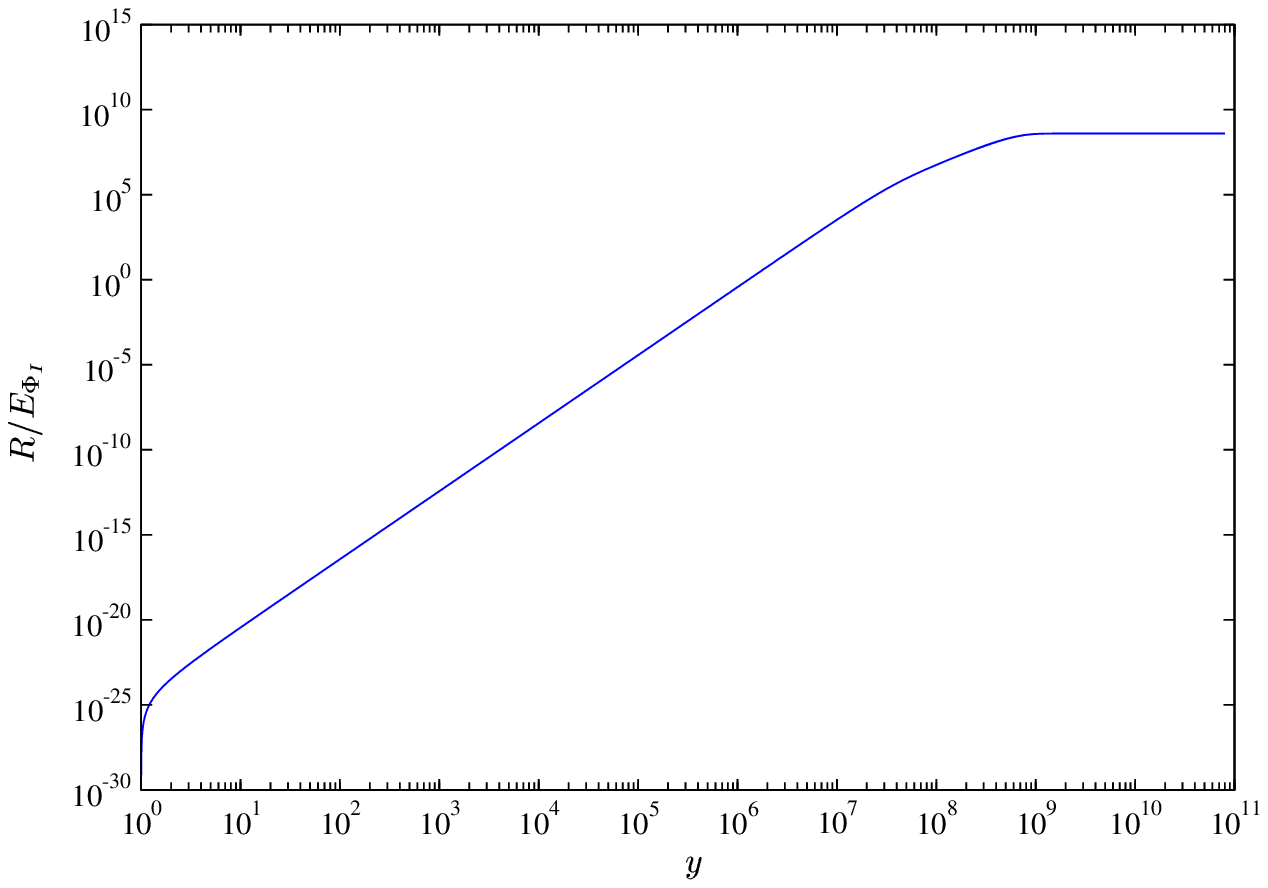}
  \label{fig:R_M19T19_K500}}
  \subfloat[]{
  \includegraphics[width=0.5\textwidth]{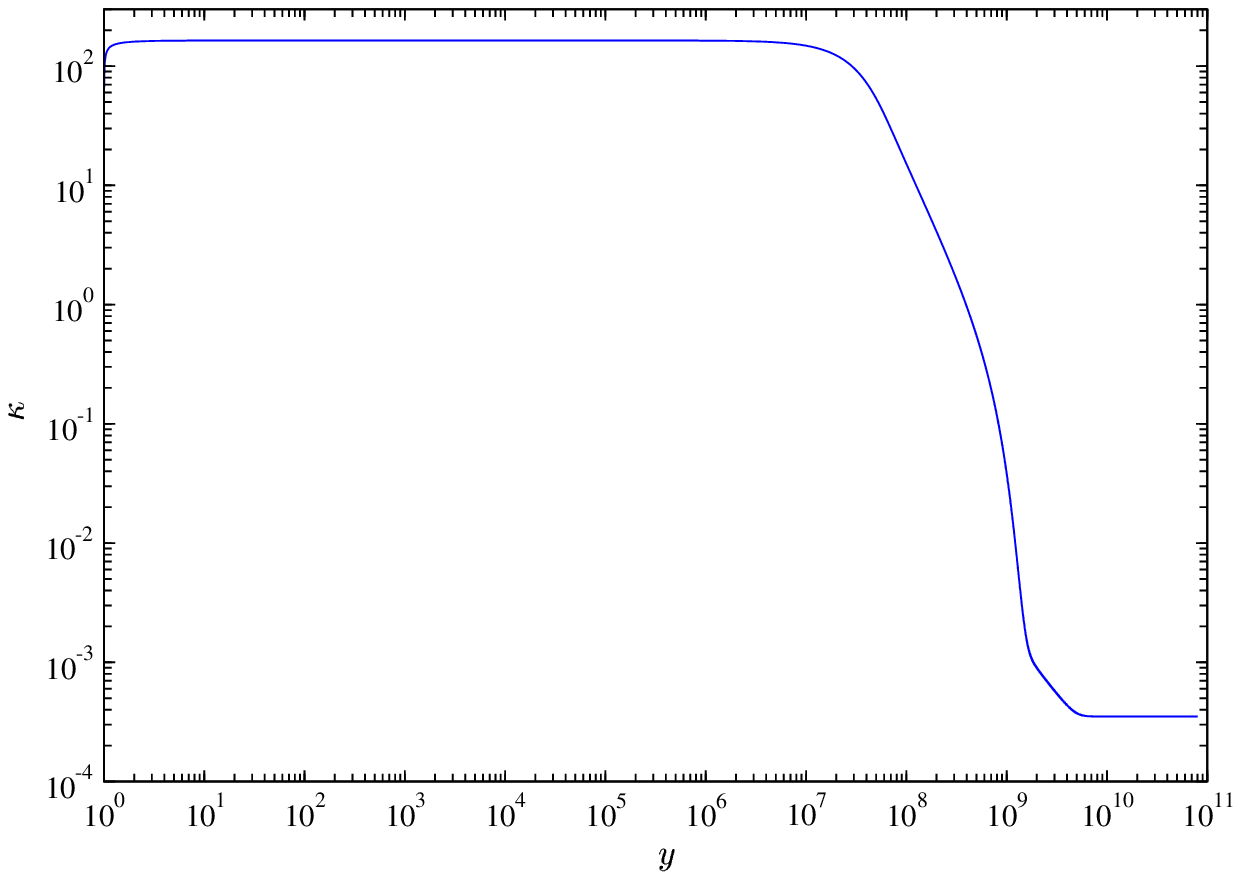}  
  \label{fig:k_M19T19_K500}}
     \subfloat[]{
  \includegraphics[width=0.5\textwidth]{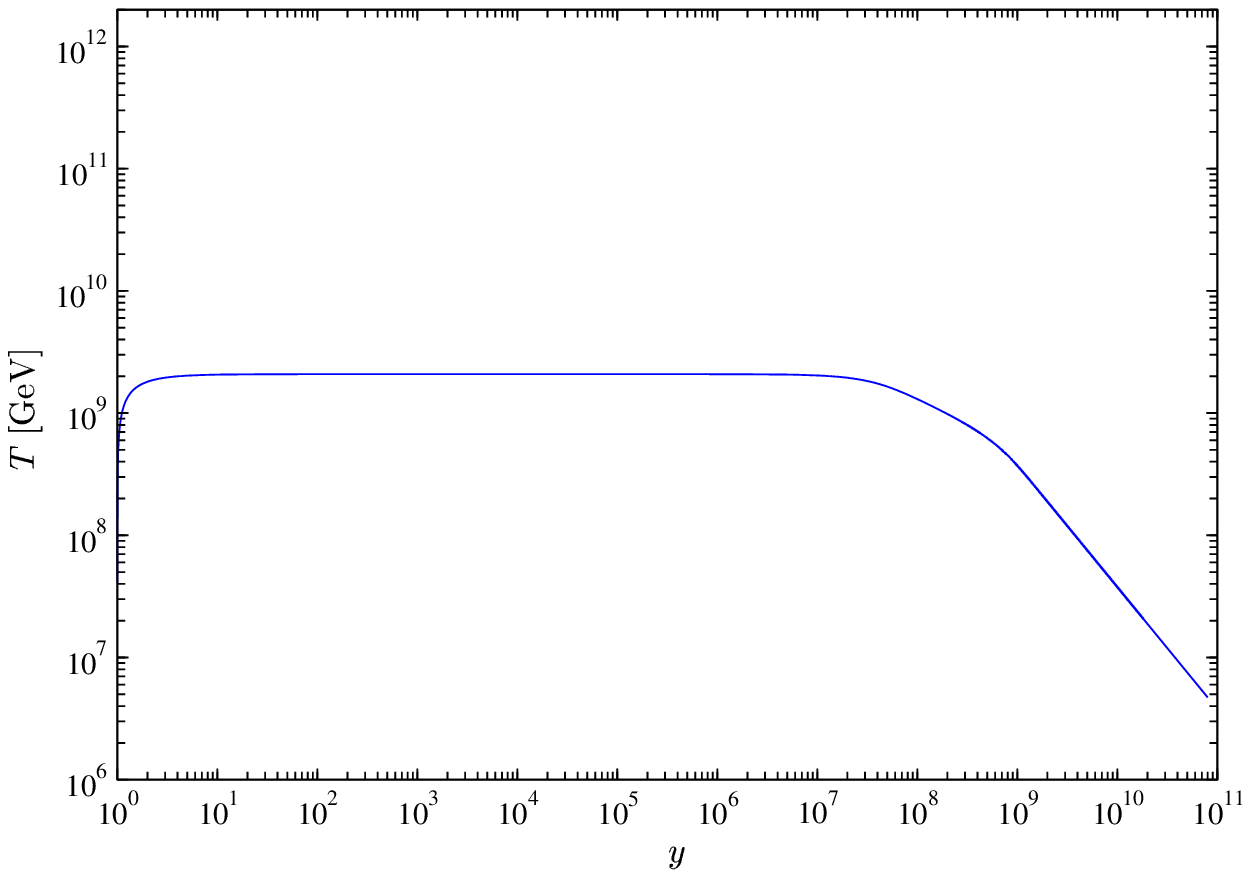}  
  \label{fig:T_M19T19_K500}}
   \subfloat[]{
  \includegraphics[width=0.5\textwidth]{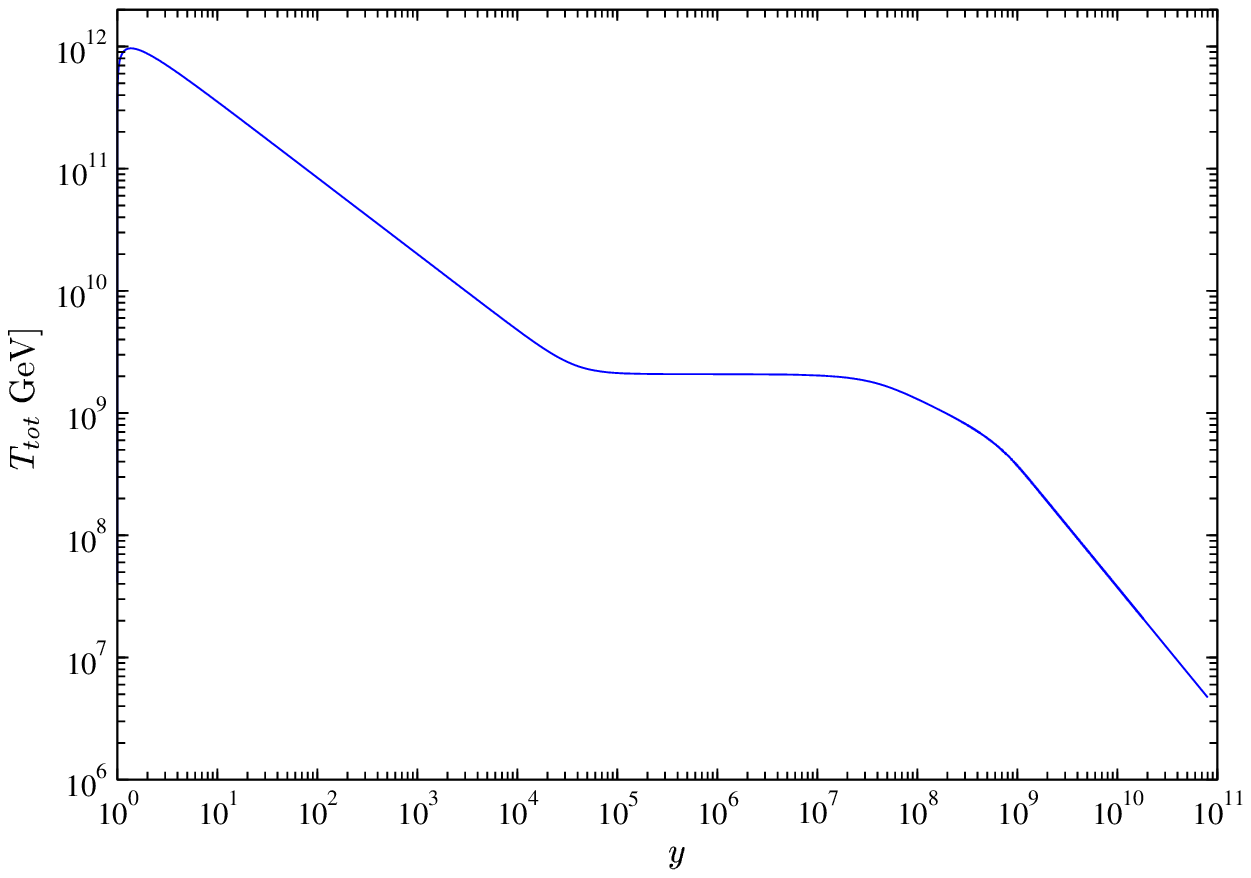} 
  \label{fig:T-tot_M19T19_K500}}
\caption{\small{The evolution of $E_{\Phi}/E_{\Phi_{I}}$,
  $E_{N}/E_{\Phi_{I}}$, $R$, $\kappa$, $T$ and $T_{tot}$is shown for
  $T_{RH}=10^{9}$ GeV at $K=500$}}
\end{figure}

As an example, let us again consider the case $T_{RH}=10^9\,$GeV but now with
$K=500$ in some detail. As one can see in figure~\ref{fig:Fi_M19T19_K500}, the
inflaton again starts to decay at $y \sim 10^{9}$ and has decayed completely
at about $y \sim 10^{10}$.  The rescaled right-handed neutrino energy density,
cf.\ figure~\ref{fig:N_M19T19_K500}, becomes constant already at $y \sim 3
\times 10^{7}$ and, due to the larger value of $K$, the $N_{1}$ start to decay
much earlier than in the weak washout, already at $y \sim 10^{9}$.  Because of
inverse decay processes, which in thermal equilibrium balance the decay
processes, the decrease of the neutrino abundance is much slower than in the
previous example.  The neutrinos have fully decayed at $y = 5 \times 10^{10}$,
i.e.\ somewhat later than in the weak washout regime. The transition to a
radiation dominated universe occurs at $y \sim 10^9$ as can be seen in
figure~\ref{fig:R_M19T19_K500}.  At this point the rescaled radiation energy
density $R/E_{\Phi}$ becomes constant. The stronger interactions of the heavy
right-handed neutrino with the SM particles have a strong impact on the
efficiency factor.  It rises quickly to $\sim 160$ and remains constant for a
long time.  When the energy of the right-handed neutrino becomes constant at
$y \sim 3 \times 10^{7}$ the efficiency factor decreases. This decrease even
accelerates when the neutrinos start decaying, and once the neutrinos have
entirely decayed away, the final efficiency factor reads $\kappa_{f} \sim 4
\times 10^{-3}$.  In figure~\ref{fig:T_M19T19_K500} one can see that the
temperature of the thermal bath of standard model particles rises quickly and
then remains constant at $T \sim 3 \times 10^{9}$.  When $E_{N}/E_{\Phi_{I}}$
becomes constant the temperature starts to decrease slowly and becomes
inversely proportional to the scale factor once the universe is radiation
dominated. The physical reheating temperature is now \mbox{$T_{RH}^{N} \sim
  10^{9} \GeV$}, approximately one order of magnitude larger than in the weak
washout example we had considered previously.

\subsection{Behaviour for $T_{RH} \ll M_{1}$}

The behaviour of the final efficiency factor for reheating temperatures
$T_{RH} \ll M_{1}$ is somewhat different than for the values of
$T_{RH}$ discussed above. As an example, let us discuss the case 
$T_{RH} \sim 10^{8}\,$GeV.



As we can see in figure~\ref{fig:non_thermal_kf_M9Tall}, $\kappa_{f}$
is now almost independent of $K$ and of order 10. Only in the limit $K
\rightarrow 0$ does one obtain the same value for $\kappa_{f}$ as for
$T_{RH} \gtrsim M_{1}$. For larger values of $K$ we see that washout
is now completely negligible and that $\kappa_{f}$ remains almost
constant even for very large $K$, where one only observes a small
decrease of $\kappa_{f}$. The effect of entropy production in $N_1$
decays is also somewhat weaker now. Indeed, for such a low reheating
temperature the decay width of the inflaton is much smaller than the
decay width of the lightest right-handed neutrino, $\Gamma_{\Phi}
\ll\Gamma_{N}$. Hence, the $N_{1}$'s always decay strongly
out-of-equilibrium and instantaneously after having been produced in
inflaton decays. Therefore, the physical reheating temperature
$T_{RH}^{N}$ becomes nearly independent of $K$ and is given directly
by $T_{RH}$ since the time period of a neutrino dominated universe is
negligibly short. For even lower reheating temperatures one expects
neither an effect due to washout for $K>1$ nor due to entropy
production for $K<1$ since the right-handed neutrino decay again
follows instantaneously the inflaton decay. This is the scenario
sketched at the beginning of section~\ref{sec:lg_i_decay} which had
been considered in the literature before.

\subsection{Dependence of the Results on $M_{\Phi}$ and $M_{1}$}

An obvious question is how the results presented so far depend
on the masses of the inflaton, $M_{\Phi}$, and the right-handed
neutrino, $M_1$.
%
A change of the inflaton mass $M_{\Phi}$ can also be parametrized
by a variation of the reheating temperature, cf.\ Eq.~(\ref{eq:4}).
Hence, for our purposes it is equivalent to a change in the
inflaton-neutrino coupling $\gamma$, which we have already discussed
above. The only limit is set by the
kinematical lower bound on the inflaton mass, $M_{\Phi}>2M_{1}$.

\begin{figure}[t]
  \centering
  \includegraphics[width=0.7\textwidth]{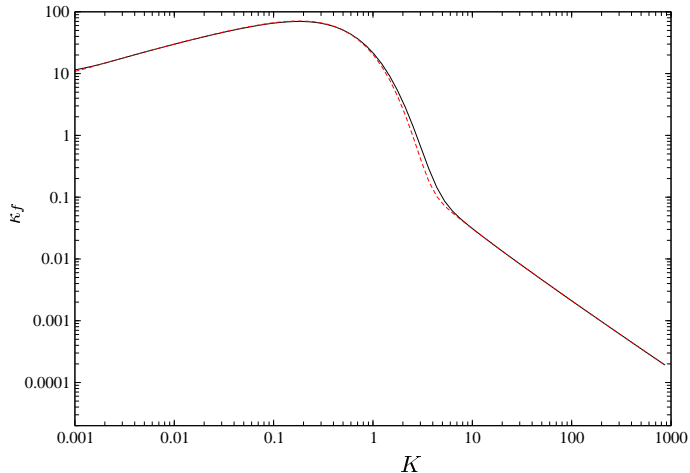}
  \caption{\small{$\kappa_{f}$ for $M_{1}=10^{9}\,$ GeV (solid) and
      $M_{1}=10^{7}\,$ GeV (dashed)}}
  \label{fig:kf_diff}
\end{figure}

As an example for the dependence of the results on $M_1$, we have
plotted the final efficiency factor $\kappa_f$ for $M_1=10^9\,$GeV
and $M_1=10^7\,$GeV in fig.~\ref{fig:kf_diff}. As one can see, the
differences between the two cases are negligible.
%
This is due to the fact that
the right-handed neutrinos always decay when they are non-relativistic. For
$M_{1}=10^{9}$ GeV this happens instantaneously after the production
in inflaton decays. For $M_{1}=10^{7}$ GeV, on the other hand, the
$N_{1}$ abundance remains constant until the universe has cooled down
to a temperature $\sim 10^{7}$ GeV. Then the non-relativistic
neutrinos decay in the same way as for $M_{1}=10^{9}$ GeV. A variation
of $M_{1}$ changes the point in time of the decay process and not the
process itself and hence $\kappa_{f}$ is unchanged. This is completely
analogous to the situation in thermal leptogenesis, where $\kappa_f$
is also independent of $M_1$, as long as $M_1\lesssim10^{13}\,$GeV.


\subsection{Lower Bound on $T_{RH}^{N}$ and $M_{1}$}

Finally, let us discuss
the impact a dominant initial neutrino abundance
has on the lower limits on the physical reheating temperature and
the right-handed neutrino mass. Demanding succesful leptogenesis
in the standard thermal case leads to the lower limit 
\cite{Buchmuller:2004nz}
\begin{equation}
\label{eq:23}
  M_{1} > M^{min}_{1}(K)
  \approx 6.4 \times 10^{8} \textrm{GeV}
  \left(\frac{\eta^{CMB}_{B}}{6 \times 10^{10}} \right) \left(\frac{
      0.05\, \textrm{eV}}{m_{\textrm{atm}}} \right) \kappa^{-1}_{f}(K)\;.
\end{equation}
Hence, assuming a thermal initial abundance of right-handed neutrinos and
in the limit $K\to0$ the absolute lower limits on $M_1$ and the reheating
temperature read
\begin{equation}
\label{eq:24}
  T_{RH}, \; M_{1} \gtrsim 4 \times 10^{8}\,\textrm{GeV}\;.
\end{equation}
This was obtained at $3\sigma$ using
$\eta_{B}^{\textrm{CMB}}=(6.3 \pm 0.3) \times 10^{-10}$ for the baryon
asymmetry and $\Delta m_{\textrm{atm}}^{2} =(1.2-4.8) \times 10^{-3}
\eV^{2}$ for the mass square difference in atmospheric
neutrino oscillations. 

In the case of a zero initial neutrino abundance in thermal leptogenesis,
the lower limits are reached at $K\simeq1$ and read
\begin{equation}
  T_{RH},\;M_1\gtrsim 2\times10^9\,\textrm{GeV}\;.
\end{equation}

\begin{figure}[t]
  \centering
  \includegraphics[width=0.6\textwidth]{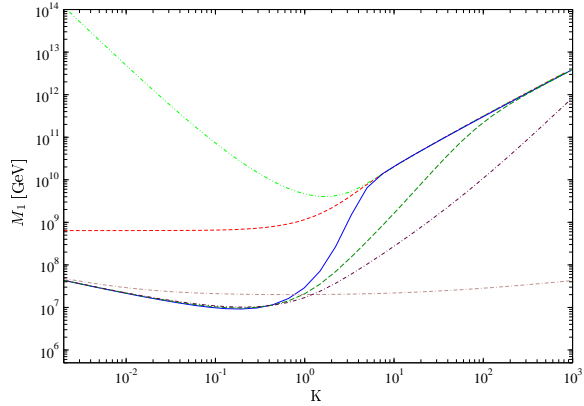}
  \caption{\small {The lower bound on $M_{1}$ is shown for thermal
    (short-dashed), zero (point-point-dashed) and dominant initial
    $N_{1}$ abundance for $T_{RH}=10^{9}$ GeV (solid), $T_{RH}=5
    \times 10^{8}$ Gev (long-dashed), $T_{RH}=3.75 \times 10^{8}$ GeV
    (point-dashed and $T_{RH}=10^{8}$ GeV (point-dashed-dashed))}}
    \label{fig:bound_on_M_all}.
\end{figure}

In our case, the final efficiency factor in the limit $K\to0$ is
greatly enhanced, i.e.\ the lower limit $M_1^{\textrm{min}}$ gets
relaxed accordingly. It will now not only depend on $K$, as in
thermal leptogenesis, but also on the reheating temperature $T_{RH}$
which parametrizes the inflaton-neutrino coupling. The results
are summarized in Fig.\ref{fig:bound_on_M_all}, where we have
also shown the lower limits from thermal
leptogenesis~\cite{Buchmuller:2004nz} for comparison.

A lower bound on $M_{1}$ again corresponds to a lower bound on the
physical reheating temperature $(T_{RH}^{N})^{\textrm{min}}$.
Note that, since $(T_{RH}^{N})^{\textrm{min}}(K) =
M_{1}^{min}(K)\, \sqrt{K}$, the lowest value of
$T_{RH}^{N}$ is achieved in the limit $\tilde{m}_{1} \rightarrow 0$
where one obtains
\begin{equation}
  \label{eq:25}
  (T_{RH}^{N})^{\textrm{min}} \sim 2.4 \times 10^{6} \GeV\;,
\end{equation}
which is two orders of magnitude lower than in the case of
thermal leptogenesis.

The neutrino oscillation data favour the effective neutrino mass, i.e.\
$\tilde{m}_1=K m_{\ast}$, to lie in the neutrino mass window 
$m_{\textrm{sol}} < \tilde{m}_1 < m_{\textrm{atm}}$, in the strong washout
regime. In this range and for large reheating temperatures,
$T_{RH}\gtrsim M_1$, we obtain the same lower limit as in thermal
leptogenesis,
\begin{equation}
  \label{eq:14}
    (T_{RH}^{N})^{\textrm{min}} \sim 
    (4 \times 10^{9} -2 \times 10^{10}) \GeV\;.
\end{equation}
For lower reheating temperatures, e.g.\ $T_{RH}= 5\times 10^{8} \GeV$, 
one can see from figure~\ref{fig:bound_on_M_all} that the bound on $M_{1}$ 
is about one order of magnitude lower than in thermal leptogenesis, hence
the lower limit on the physical reheating temperature now reads
\begin{equation}
\label{eq:26}
   (T_{RH}^{N})^{\textrm{min}} \sim (4 \times 10^{8} - 2 \times 10^{9}) \GeV.
\end{equation}



Further lowering the reheating temperature, i.e.\ the inflaton-neutrino
coupling, to $T_{RH}=10^8\,$GeV leads to an even weaker lower limit on the
right-handed neutrino mass. As already discussed for the final efficiency
factor, for such low reheating temperatures the results are almost
independent of $K$. In the limit $K\to0$ we recover for the physical
reheating temperature the result obtained above,
cf.\ Eq.~(\ref{eq:25}).

In the more interesting strong washout regime favoured by neutrino
oscillation data, the lower limit on the heavy neutrino mass reads,
\begin{equation}
  \label{eq:27}
  M_{1} \gtrsim 4 \times 10^{7} \GeV\;.
\end{equation}
 Correspondingly, the lower bound on the physical reheating temperature
gets relaxed to
\begin{equation}
\label{eq:28}
   (T_{RH}^{N})^{\textrm{min}} \sim  10^{7} \GeV\;.
\end{equation}
Hence, in the phenomenologically most interesting strong washout regime
a dominant initial neutrino abundance produced in inflaton decays can
be used to relax the rather stringent lower limit on the physical
reheating temperature obtained in thermal leptogenesis by up to three
orders of magnitude, provided the inflaton-neutrino coupling is small.

\section{Conclusions}
\label{sec:consclsions}

In this paper we have studied some aspects of non-thermal leptogenesis
as an alternative to the standard thermal leptogenesis scenario. In
particular, we investigated the interplay between inflation and leptogenesis
by considering a decay chain where the inflaton first exclusively
decays into heavy right-handed neutrinos which then decay into standard
model lepton and Higgs doublets, thereby reheating the universe and
creating the baryon asymmetry of the universe.

We have performed a full numerical study by means of a set of Boltzmann
equations and have discussed the dependence of the final efficiency
factor, corresponding to the maximal baryon asymmetry which can be
produced, on the inflaton-neutrino coupling and the heavy neutrino
Yukawa coupling. To that end, we have parametrized the inflaton-neutrino
coupling in terms of the reheating temperature $T_{RH}$ defined in the 
standard way, which, however, should not be confused with the physical
reheating temperature, since in our scenario the universe becomes radiation
dominated once the heavy neutrinos and not the inflaton have decayed.

We have mainly discussed values of $T_{RH} \sim M_{1}$. This is in
contrast to most scenarios considered before in the literature where 
$M_{1} \gg T_{RH}$ is usually assumed.
For those values the final efficiency factor is enlarged by a
factor $\sim 10-100$ in the weak washout regime compared to the one
obtained in thermal leptogenesis. In the strong washout
regime, on the other hand, the final efficiency factor that one gets
in thermal leptogenesis is reproduced, if $T_{RH}\gtrsim M_1$.
Furthermore, we have seen that for $T_{RH} \ll M_{1}$ the right-handed
neutrinos decay completely out-of-equilibrium and hence the final 
efficiency factor is almost independent of $K$, which parametrizes
the neutrino Yukawa coupling. For such reheating temperatures the 
final efficiency factor is a factor $\sim 10$ larger than in thermal 
leptogenesis. 

Increasing the efficiency of leptogenesis is particularly interesting
in light of the rather stringent upper limits on the reheating
temperature of $10^{6-7}\;$GeV obtained in certain supersymmetric
scenarios. Indeed, such an upper limit is in conflict with the lower
limit on the reheating temperature of $4\times10^8\;$GeV obtained
in thermal leptogenesis.

Here, we could show that in the weak washout regime reheating
temperatures as low as $\sim10^6\;$GeV are permissible in our
non-thermal scenario, independently of the neutrino-inflaton
coupling. In the phenomenologically more interesting neutrino
mass window in the strong washout regime, reheating temperatures
as low as $\sim10^7\;$GeV still allow for succesful leptogenesis, as
long as $T_{RH}\ll M_1$, i.e.\ as long as the neutrino-inflaton
coupling is small.



\section*{Acknowledgments}
We would like to thank J.~Pradler and G.~Raffelt for useful discussions
and suggestions. Further, we are indebted to P.~Di Bari for collaboration
during early stages of this work as well as useful comments on the
manuscript.

\appendix

\section{Variable Transformation}
\label{sec:var-trans}
When solving the Boltzmann equations it is convenient to use variables
in which the expansion of the universe has been scaled out. In analogy
to the procedure presented in~\cite{Chung:1998rq}\footnote{Note that 
$\tilde{N}_{B-L}$ is the particle density per comoving volume element for
the asymmetry.}, we shall use the following variables:
\begin{equation}
  \label{eq:10}
   \begin{aligned}
    E_{\Phi} &= \rho_{\Phi}a^{3}\;, \\
    E_{N} &= \rho_{N}a^{3}\;,\\
    \tilde{N}_{B-L} &= n_{B-L}a^{3}\;, \\
    R &= \rho_{R}a^{4}\;,
  \end{aligned}
\end{equation}
where $a$ is the scale factor of the universe. Moreover, it is convenient 
to write the Boltzmann equations as functions of the scale factor rather
than time. More precisely, we shall use the ratio of the scale factor to
its initial value,
\begin{equation}
  \label{eq:11}
    y= \frac{a}{a_{I}}\;,
\end{equation}
as time variable. For definiteness, we shall use $a_I=1$. Then the
expansion rate reads:
\begin{equation}
  \label{eq:13}
  H= \sqrt{ \frac{8 \pi (a_{I} E_{\Phi} y+ a_{I}E_{N}y+R ) }
                 { 3M_{Pl}^{2}a_{I}^{4}y^{4}}}\;.
\end{equation}
Further, instead of the temperature $T$ we use the inverse temperature in
units of the heavy neutrino mass,
\begin{equation}
  \label{14}
  z=\frac{M_{1}}{T} =
  M_{1}a_{I}\left[\frac{\pi^{2}g^{\ast}}{30R}\right]^{\frac{1}{4}}\,y\;.
\end{equation}
Then, the rescaled equilibrium energy density of $N$ can be expressed as:
\begin{equation}
  \label{eq:15}
  E_{N}^{eq}= \rho_{N}^{eq}a^{3} = \rho_{N}^{eq}a_{I}^{3}y^{3}=
  \frac{a_{I}^{3}M_{1}^{4}y^{3}}{\pi^{2}}\,
  \left[\frac{3}{z^{2}}K_{2}(z)+\frac{1}{z}K_{1}(z)\right]\;.
\end{equation}
 In terms of these rescaled variables the Boltzmann equations, cf.\
Eqs.~(\ref{eq:9}), are given by:
\begin{equation}
  \label{eq:16}
  \begin{aligned}
  \frac{dE_{\Phi}}{dy} &=  \frac{\Gamma_{\Phi}}{H} \frac{E_{\Phi}}{y}\;, 
  \\ 
  \frac{dE_{N}}{dy} &= \frac{\Gamma_{\Phi}}{H} \frac{E_{\Phi}}{y}
  - \frac{\Gamma_{N}}{Hy} \left( E_{N} - E_{N}^{eq}
  \right)\;,
  \\ 
  \frac{d\tilde{N}_{B-L}}{dy} &= - \frac{\Gamma_{N}}{Hy} \left[ \epsilon_{1}
    \left(\tilde{N} -
      \tilde{N}^{eq} \right) + \frac{n_{N}^{eq}}{n_{R}^{eq}} \tilde{N}_{B-L}
  \right]\;,
  \\ 
  \frac{dR}{dy} &= \frac{ \Gamma_{N} a_{I}}{H} \left( E_{N} -
   E_{N}^{eq} \right)\;.
  \end{aligned}
\end{equation}


As already mentioned in the main text, we fix the mass of right-handed
neutrino and inflaton to $10^9\;$GeV and $10^{13}\;$GeV, respectively.
The reheating temperature $T_{RH}$ is used to parametrize the 
inflaton-neutrino coupling, and $K$ parametrizes the Yukawa coupling
of the heavy nuetrinos.

For the initial energy density of the inflaton or the universe's energy
density we have from the condition $\Gamma_{\Phi}=H(a_{I})$:
\begin{equation}
\label{eq:17}
\rho_{I}=\frac{3}{8\pi}M_{\Phi}^{2}M_{Pl}^{2}\;.
\end{equation}
Note that $N_{B-L}$ used in~\cite{Buchmuller:2004nz} is related to
$\tilde{N}_{B-L}$, defined in Eq.~(\ref{eq:16}), by the following
relation:
\begin{equation}
  \label{eq:19}
  N_{B-L} = \frac{n_{B-L}}{n_{\gamma}} = 
  \left[ \frac{\pi^{4}}{30\zeta(3)} \right]
  \frac{n_{B-L}}{\rho_{\gamma}} \, T =
  \left[ \frac{\pi^{
4} {g^{\ast}}^{\frac{3}{4}}}{30^{3}
      {\zeta(3)}^{4}} \right]
  R^{-\frac{3}{4}}\, \tilde{N}_{B-L}\;,
\end{equation}
Defining the final efficiency factor as~\cite{Buchmuller:2004nz}
\begin{equation}
  \label{eq:20}
  \kappa_{f} = - \frac{4}{3} \epsilon^{-1} N_{B-L}\;,
\end{equation}
we can compare the results for dominant initial $N_1$
abundance with those obtained for thermal and zero initial abundance
in previous calculations~\cite{Buchmuller:2004nz}.

\end{document}